\documentclass[fleqn,usenatbib]{mnras}
\usepackage{newtxtext,newtxmath}
\usepackage{longtable}
\usepackage{verbatim, longtable}	
\usepackage[figuresright]{rotating}

\usepackage[T1]{fontenc}
\usepackage{CJKutf8}

\DeclareRobustCommand{\VAN}[3]{#2}
\let\VANthebibliography\thebibliography
\def\thebibliography{\DeclareRobustCommand{\VAN}[3]{##3}\VANthebibliography}

\usepackage{graphicx}	% Including figure files
\usepackage{amsmath}	% Advanced maths commands
\title[variability of AGNs]{A universal relationship between the variability timescale and black hole mass in black hole jetted and non-jetted accreting systems}

\author[Yongyun Chen et al.]{Yongyun Chen\begin{CJK*}{UTF8}{gkai}(陈永云)\end{CJK*}\thanks{E-mail: ynkmcyy@yeah.net}$^{1}$,
Qiusheng Gu\begin{CJK*}{UTF8}{gkai}(顾秋生)\end{CJK*}\thanks{E-mail: qsgu@nju.edu.cn}$^{2}$,
Junhui Fan\begin{CJK*}{UTF8}{gkai}(樊军辉)\end{CJK*}$^{3}$,
\and Dingrong Xiong\begin{CJK*}{UTF8}{gkai}(熊定荣)\end{CJK*}$^{6}$
 Xiaoling Yu \begin{CJK*}{UTF8}{gkai}(俞效龄)\end{CJK*}$^{1}$,
Xiaogu Zhong \begin{CJK*}{UTF8}{gkai}(钟晓谷)\end{CJK*}$^{1}$,
\and Xiaotong Guo \begin{CJK*}{UTF8}{gkai}(郭晓通)\end{CJK*}$^{5}$,
\\
$^{1}$College of Physics and Electronic Engineering, Qujing Normal
University, Qujing 655011, P.R. China\\
$^{2}$School of Astronomy and Space Science, Nanjing University, Nanjing 210093, P. R. China\\
$^{3}$Center for Astrophysics,Guang zhou University,Guang zhou510006, China\\
$^{4}$Yunnan Observatories, Chinese Academy of Sciences, Kunming 650011, China\\
$^{5}$School of mathematics and physics, Anqing Normal University 246011, P. R. China\\
}

\date{Accepted XXX. Received YYY; in original form ZZZ}

\pubyear{2026}

\begin{document}
\label{firstpage}
\pagerange{\pageref{firstpage}--\pageref{lastpage}}
\maketitle

\begin{abstract}
A long-term variability study spanning a range of black hole mass systems, from microquasars hosting stellar-mass black holes to active galactic nuclei (AGNs) harboring supermassive black holes, provides new insights into the physics of relativistic jets. In this work, we investigate the optical variability of both jetted and nonjetted AGNs. We apply a stochastic process known as the Damped Random Walk (DRW) to model light curves from the Zwicky Transient Facility (ZTF) DR23. Our results show that the mass-scaled characteristic timescales across the black hole mass exhibit a linear relationship with a slope of $\sim$0.35–0.50. This analysis confirms a previously observed correlation between the damping timescales and black hole mass and extends it by incorporating 125 newly identified non-jetted AGNs selected from the Burst Alert Telescope (BAT) AGN catalogue. The derived slope of the relation between the damping timescales and black hole mass aligns with recent theoretical predictions, supporting the presence of a universal accretion mechanism in AGNs across different mass scales. Furthermore, our findings suggest a novel implication: the properties and production mechanisms of relativistic jets may be largely independent of black hole mass.
\end{abstract}


%%%%%%%%%%%%%%%%%%%%%%%%%%%%%%%%%%%%%%%%%%%%%%%%%%

%%%%%%%%%%%%%%%%% BODY OF PAPER %%%%%%%%%%%%%%%%%%

\section{Introduction}
Jet structures have been observed across a wide range of black hole accretion systems, from stellar-mass black holes (SBHs) to supermassive black holes (SMBHs), exhibiting remarkable diversity in scale and environment. Notable examples include microquasars—stellar-mass black hole binary systems—and blazars, a subclass of active galactic nuclei (AGN) with relativistic jets closely aligned to our line of sight \citep{Mirabel1999, Urry1995, Padovani2017, Blandford2019}. The powering of relativistic jets in these systems is widely attributed to two primary mechanisms: energy extraction from a rotating black hole and/or the rotational energy of the accretion flow mediated by large-scale magnetic fields \citep{Blandford1977, Blandford1982}. Subsequent theoretical advancements have refined these models, offering deeper insights into the underlying physical processes \citep{Williams1995, Williams2004, Zamaninasab2014, Bellan2018}. While the exact mechanism responsible for jet launching remains incompletely understood, these investigations consistently emphasize the central role of accretion disks in jet formation.

Both direct and indirect observations provide strong evidence for a physical connection between accretion processes and jet formation. In the context of supermassive black holes, numerous studies have revealed a significant correlation between the kinetic power of relativistic jets and the luminosity of the accretion disk. For example, \cite{Rawlings1991} were among the first to report this correlation, which has since been extensively examined and robustly confirmed by a series of follow-up investigations \citep[e.g.,][]{Cao1999, Ghisellini2009, Ghisellini2010, Sbarrato2014, Ghisellini2014, Chen2015, Chen2023a, Chen2023b, Chen2023c, Chen2024}. These results indicate that the energy available for jet launching is closely tied to both the rate at which matter accretes onto the black hole and the efficiency of the underlying accretion process.

In microquasars, the X-ray spectra typically exhibit transitions between hard and soft states \citep{Belloni2000, Zdziarski2002, Done2007}. The hard state is characterized by a dominant power-law component in the X-ray spectrum, whereas the soft state is dominated by a thermal component originating from the accretion disk. Notably, many microquasars show radio jet emissions, and a significant correlation between radio and X-ray fluxes has been observed during the hard state \citep{Markoff2001, Corbel2003, Markoff2005, Wilms2006, Zdziarski2020}. For instance, X-ray jet activity in the microquasar XTE J1550-564 was detected exclusively during its hard state \citep{Corbel2002}. This phenomenon may arise from two plausible mechanisms. First, energy from the accretion flow may be injected into the jet \citep{Livio2003}, with a fraction of the accretion-released energy being channeled to accelerate and heat particles within the jet. Second, the jet may originate from an extended corona serving as its base \citep{Wang2021}, where the corona—a hot, tenuous region above the accretion disk—acts as the site of initial particle acceleration leading to jet formation. Importantly, the corona is dynamically coupled to the accretion disk, where cold gas evaporates into the hot corona at low accretion rates, and hot gas condenses back into the disk at high rates, driving spectral state transitions \citep{Yuan2014}. The observed correlation among the properties of the accretion disk, corona, and jet defines what is known as the ``fundamental plane of black hole activity." This empirical relationship applies not only to stellar-mass black holes but also extends to supermassive black hole systems, as demonstrated by some studies \citep{Merloni2003, Kording2006, Nisbet2016}. These findings provide robust evidence for a universal physical connection between accretion and jet processes across a broad range of black hole masses, indicating that the underlying mechanisms of jet formation are fundamentally similar irrespective of black hole mass.

Accretion disks and jets commonly exhibit significant variability across the electromagnetic spectrum \citep[e.g.,][]{Kaastra1989, Urry1996, Kawaguchi1998, Cai2022, Tian2023}. A robust method for analyzing such variability is to derive the power spectral density (PSD) from observed light curves (LCs). The PSDs of accretion disks and jets are typically modeled as a broken power law (BPL)\footnote{That is, a power law with a slope of $\sim$2 that transitions to white noise at the characteristic damping frequency, $f_{\rm broken}$, where the corresponding timescale is defined as $\tau_{\rm damping} = 1/(2\pi f_{\rm broken})$}. The characteristic damping timescale (denoted as $\tau_{\text{damping}}$, and hereafter referred to simply as the characteristic timescale) of an accretion disk is believed to be closely tied to several fundamental physical processes: thermal processes governing the restoration of thermal equilibrium; dynamical processes associated with orbital motion; and viscous processes involving the diffusion of mass inflow \citep{Czerny1999, Czerny2006, Suberlak2021}.

Several studies have reported scaling relationships for the characteristic timescale across different black hole mass scales. \cite{McHardy2006} first identified a scaling relation (\(\tau_{\text{damping}} \propto M_{\text{BH}}\)) in the X-ray variability of non-jetted supermassive black hole systems, suggesting that the characteristic timescale of accretion disk variability scales linearly with black hole mass. In the submillimeter waveband, \cite{Bower2015} also found a  linear correlation between characteristic timescale and black hole mass for low-luminosity AGNs. Subsequently, \cite{Scaringi2015} and \cite{Burke2021} discovered a similar scaling relationship in the optical band between non-jetted SBHs and SMBHs, further supporting the existence of a common physical mechanism underlying accretion disk variability across mass scales. More recently, \cite{Zhang2024} and \cite{Sharma2025} reported an analogous scaling relation in the gamma-ray band for non-jetted SBHs and SMBHs systems. Collectively, these findings strongly suggest a universal physical origin for accretion processes in black hole systems, indicating that the fundamental mechanisms governing accretion and its associated variability are invariant across orders of magnitude in black hole mass.

\cite{Mukherjee2019} proposed that the characteristic timescales of the accretion disk and the jet may differ. However, an increasing body of evidence indicates that, in long-term LCs, the characteristic timescale of jet emission is either physically linked to or comparable in magnitude to that of the accretion disk \citep{Ruan2012, Ryan2019, Zhang2022, Zhang2023, Sharma2024}. This suggests a strong physical connection between the accretion disk and the jet, whereby variability in the disk can directly drive variability in the jet. If the disk–jet coupling is indeed significant, jetted systems across different mass scales are expected to follow a universal scaling relationship. Specifically, the characteristic timescales of both the accretion disk and the jet in SBHs, intermediate-mass black holes (IMBHs), and SMBHs should exhibit a consistent scaling pattern. However, current observational data remain limited for jetted black hole systems hosting IMBHs. IMBHs, with masses bridging those of SBHs and SMBHs, are considered a crucial missing link in the black hole mass. Their scarcity in observations hinders the identification of a unified scaling relation across all jetted systems. Recent studies have successfully identified dwarf active galactic nuclei (AGNs) harboring black holes with masses at the boundary between IMBHs and SMBHs—specifically in the range $M_{\rm BH} \sim 10^{4}-10^{6}\,M_{\odot}$—through optical spectroscopic signatures \citep[e.g.,][]{Greene2005, Reines2013, Baldassare2015, Cann2020} and variability analysis \citep[e.g.,][]{Baldassare2018, Guo2020, MartinezPalomera2020, Burke2022, Ward2022}.

In this paper, we study the relation between characteristic timescales and black hole mass from SBHs to SMBHs. The paper is organized as follows. We describe the sample and method in section 2. In section 3, we describe the results. Section 4 presents discussions. In section 5, we show the conclusions of statistical analysis. Throughout the paper, we adopt a cosmology with $H_{0}=70~\rm km~s^{-1}Mpc^{-1}$, $\Omega_{\rm m}=0.3$, and $\Omega_{\Lambda}=0.7$.

\begin{figure*}
	\includegraphics[width=16cm,height=16cm]{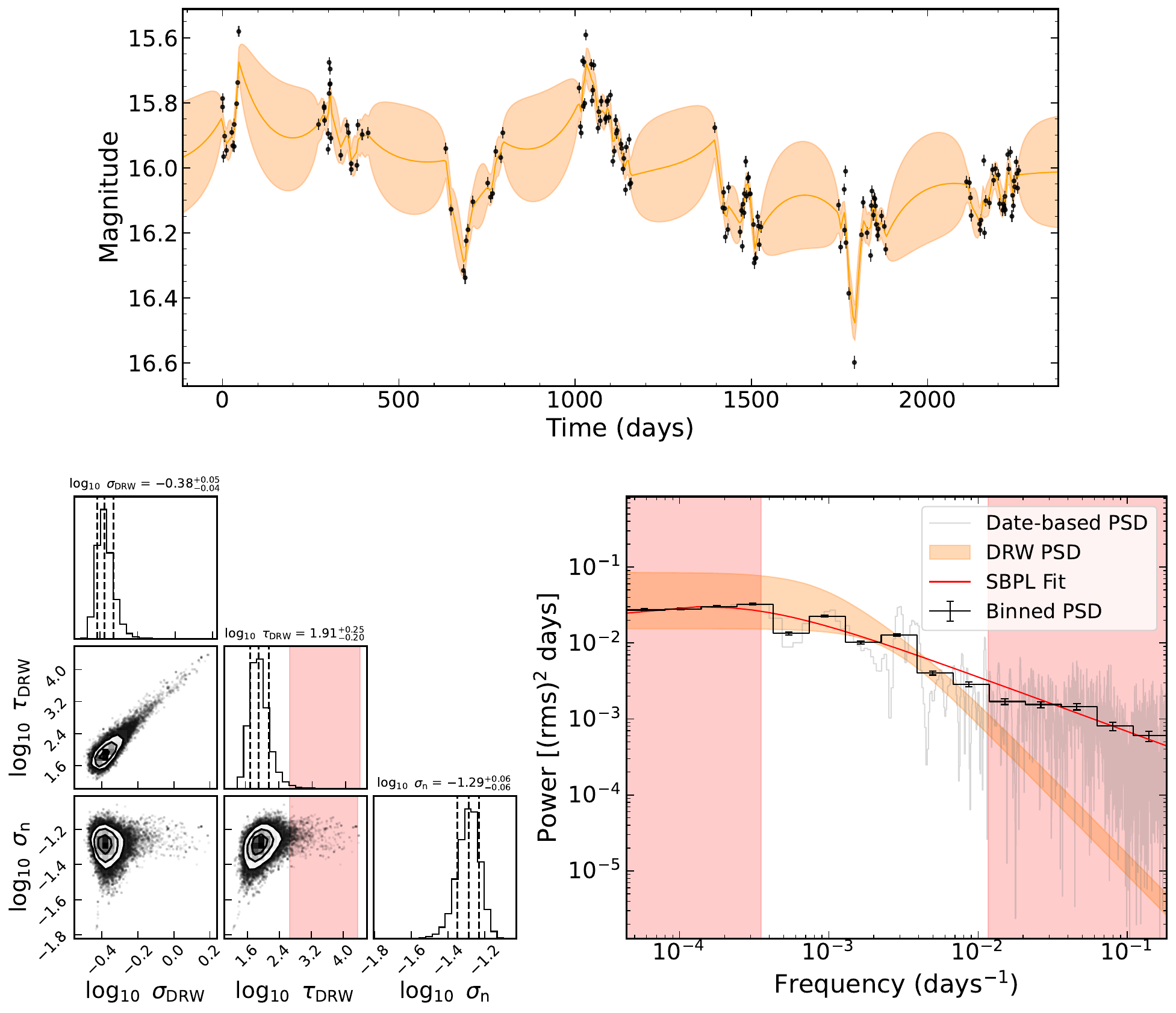}
	\centering
	\caption{
		An example of modeling the zg-band light curve of J1139.0-2323 using the damped random walk (DRW) model, implemented via the efficient Gaussian process method celerite, is presented. The upper panel shows the observed light curve together with the DRW model prediction based on the best-fit maximum likelihood parameters. The orange curve represents the optimal DRW fit, and the surrounding shaded region indicates the 1$\sigma$ confidence interval. In the lower-left subplot, the posterior distributions of the inferred DRW parameters are displayed. The lower-right panel presents both the normalized power spectral density (PSD) and its binned counterpart, including 1$\sigma$ error ranges. The orange envelope corresponds to the theoretical PSD of the DRW model, incorporating associated uncertainties. The grey line illustrates the Lomb–Scargle periodogram, while the black line shows its binned version. A broken power-law function was fitted to the binned Lomb–Scargle results, represented by the red curve. The red-shaded regions highlight time scales longer than 20\% of the total observational baseline (shown in both lower panels) and those shorter than the average sampling interval (indicated in the lower-right panel). Discrepancies between the empirical Lomb–Scargle periodogram and the modeled PSD likely arise from multiple factors: challenges in reliably estimating PSDs from unevenly sampled time series using Fourier-based methods, the influence of photometric measurement errors on spectral estimation, and potential deviations of the intrinsic variability from the assumed DRW model \citep{Stone2022}.
	}
	\label{SEDmodel}
\end{figure*}

\section{The sample and method}
\subsection{The Sample}
Our sample of jetted AGNs is drawn from \cite{Xiong2025}. Firstly, \cite{Xiong2025} got the 1684 AGNs detected by the Fermi Large Area Telescope \citep[4LAC-DR3;][]{Ajello2022}, which have reliable g-band light curves from the Zwicky Transient Facility \citep{Masci2019}. Secondly, they employed a damped random walk (DRW) model to fit the Zwicky Transient Facility (ZTF) light curves of 1684 AGNs. To obtain reliable measurements of the damping timescale, \cite{Xiong2025} used the three criteria from the work of \cite{Burke2021}: (1) the damping timescale satisfies $\tau_{\rm DRW} < 0.1 \times$ the observational baseline; (2) $\tau_{\rm DRW} >$ the mean cadence; (3) signal-to-noise criterion: $\sigma_{\rm DRW}^{2} > \sigma_{n}^{2} + \bar{dy}^{2}$. The above $\sigma_{\rm DRW}$ represents the variability amplitude, $\tau_{\rm DRW}$ is the damping timescale, $\sigma_n$ is the excess white noise amplitude, and $\bar{dy}$ denotes the mean photometric uncertainty of the light curve. They obtained 606 jetted AGNs with good model fitting met these three criteria. Finally, they crossmatched the 606 jetted AGNs with the catalog of \cite{Paliya2021} and got 183 jetted AGNs with measured black hole mass and accretion disk luminosity. From these results, \cite{Xiong2025} derived the intrinsic optical variability timescale using Doppler factor beaming correction,

 \begin{equation}
 \tau_{\rm DRW}^{\rm in} = \tau_{\rm obs}\delta/(1+z), 
 \end{equation}
where $\tau_{\rm obs}$ denotes the observed timescale, $\delta$ represents the Doppler factor, and $z$ is the redshift. \cite{Xiong2025} got the Doppler factor from the catalog of \cite{Liodakis2018}. \cite{Liodakis2018} calculated the radio‑band variability Doppler factor for the largest sample of radio‑bright blazars, using the relation connecting variability brightness temperature and intrinsic brightness temperature. When the Doppler factor cannot be obtained from the table provided by \cite{Liodakis2018}, \cite{Xiong2025} used gamma-ray luminosity to calculate the Doppler factor, $\log\delta=\log f_{b}/(-2-\alpha)$, where $\alpha$ is the spectral index of $\gamma$-ray, defined as the photon spectral index minus 1 \citep{Urry1995}. The $f_{b}$ is the beaming factor, which can be estimated for use $f_{b}=10^{-3.404}(L_{\rm obs}/10^{49})^{-0.378\pm0.052}$, and $L_{\rm obs}$ is the observed gamma-ray luminosity \citep[see Table 2 and Table 3 of \cite{Xiong2025}; ][]{Nemmen2012, Xiong2025}. The relevant data is presented in Table \ref{table00}.

We also get the microquasar sample from the work of \cite{Zhang2024}. \cite{Zhang2024} selected the 30 microquasars selected from \cite{Brocksopp2002, Choudhury2004, Wilms2006, Gupta2006, Takahashi2009, Vila2012, Dunn2010, Gallo2014, Abeysekara2018, Xie2020, Koljonen2021, Bahramian2022}. \cite{Zhang2024} employed the standard data reduction pipelines to derive the hard-state light curves, rejecting sources without XMM-Newton observations or those lacking a sufficient number of observational data points. They also used the DRW model to fit the  light curves and got the characteristic timescales of seven microquasars (see Table \ref{table0}). The reference of black hole mass is listed as follows, SS 433: \cite{Picchi2020}; GRO J1655-40: \cite{Motta2014}; IGR J17091-3624: \cite{Iyer2015}; LS I+61 303: \cite{Zabalza2011}; GRS 1915+105: \cite{Hurley2013}; H1743-322: \cite{Molla2016}; and Cygnus X-1: \cite{Orosz2011}.

Our sample of non-jetted AGNs is drawn from the hard X-ray all-sky survey conducted by the Burst Alert Telescope AGN Spectroscopic Survey \citep[BASS DR2]{Koss2022, Oh2022}. Their catalog includes 858 AGNs with measured black hole masses, accretion disk luminosity and redshifts. We extract the g-band light curves for 545 out of these 858 AGNs from the DR23 public data release of the Zwicky Transient Facility (ZTF)\footnote{https://irsa.ipac.caltech.edu/data/ZTF/docs/releases/dr23} \citep{Masci2019}. We use only the g-band light curves because the g-band magnitude is less contaminated by host galaxy starlight compared to the r-band, thereby providing a cleaner probe of nuclear variability. To ensure the reliability of the light curves, we adopt the methodology outlined in \cite{Negi2022}. The selection criteria are as follows: (1) Each source must have more than 30 observational data points; (2) For sources associated with multiple observation IDs, we select only the light curve corresponding to the ID with the largest number of epochs, as merging light curves from different IDs may introduce artificial variability \citep{vanRoestel2021}; (3) Only data with a catflags value of 0 from ZTF DR23 are included, indicating observations obtained under optimal observing conditions; (4) Light curves that lack sufficient signal-to-noise ratio or temporal baseline to reliably constrain the damping timescale are excluded \citep{Burke2021}. After applying these criteria, 495 of the initial 545 AGNs remain in the final sample.

\subsection{The DRW Model}
Modeling active galactic nuclei (AGN) light curves in the time domain using damped random walk (DRW) models has become increasingly prevalent. This approach has been widely adopted in numerous studies \citep[e.g.,][]{Kelly2009, Kelly2014, Simm2016, Suberlak2021, Zhang2022, Stone2022, Burke2021, Burke2022, Wang2023, Zhang2023, Zhang2024, Chen2025}. The DRW method effectively mitigates issues inherent in frequency-domain analysis—such as aliasing and spectral leakage—caused by irregular sampling and windowing effects, making it particularly well suited for analyzing AGN light curves. The DRW model represents the simplest case within a broader class of continuous autoregressive moving average (CARMA) models for Gaussian processes \citep[e.g.,][]{Kelly2014, Burke2021}. \cite{Burke2021} conducted a comprehensive evaluation of the advantages and limitations of the DRW model relative to higher-order CARMA models. In this work, we employ Gaussian process regression to fit a DRW model to the observed light curves, implemented using the \texttt{celerite} package \citep{Foreman-Mackey2017}. A Gaussian process (GP) is fully characterized by its covariance function, commonly referred to as the kernel. The kernel function of the DRW model, including an added white noise term in \texttt{celerite}, can be expressed as follows:

\begin{equation}
	k(t_{nm})=2\sigma_{\rm DRW}^{2}{\rm exp}(-t_{nm}/\tau_{\rm DRW})+\sigma_{n}^{2}\delta_{nm}, 
\end{equation}
where $t_{nm} = |t_n - t_m|$ denotes the time lag between measurements $n$ and $m$, and $\delta_{nm}$ is the Kronecker delta function. This formulation is mathematically equivalent to the definition of the structure function (SF),

\begin{align}
	{\rm{SF}}^2 & = \rm{SF}_{\infty}^2(1 - {\rm{ACF}}(t_{nm})) \nonumber \\
	& = 2 \sigma_{\rm{DRW}}^2 (1 - \exp{(\rm -t_{nm} / \tau_{\rm{DRW}}}))\ , 
	\label{eq:drw}
\end{align}
where the structure function at infinity is given by $\rm{SF}_{\infty} = \sqrt{2} \sigma_{\rm{DRW}}$, and the autocorrelation function is defined as ${\rm{ACF}}(t_{nm}) = \exp{(-t_{nm} / \tau_{\rm{DRW}})}$. We apply the DRW model to fit the light curves of 495 AGNs using the {\it taufit}\footnote{https://github.com/burke86/taufit} code \citep{Burke2021}. An example of the fitting result is shown in Figure \ref{SEDmodel}. Following \cite{Burke2021}, we also impose the following minimal quality criteria on the light curves for our final sample: (1) the damping timescale satisfies $\tau_{\rm DRW} < 0.1 \times$ the observational baseline; (2) $\tau_{\rm DRW} >$ the mean cadence; (3) signal-to-noise criterion: $\sigma_{\rm DRW}^{2} > \sigma_{n}^{2} + \bar{dy}^{2}$. After applying these cuts, reliable damping timescales are obtained for 132 out of the 495 sources. Additionally, we select only radio-quiet AGNs (i.e., non-jetted AGNs) based on the radio ($L_{\rm R}$) to X-ray ($L_{\rm X}$) luminosity ratio \citep{Terashima2003, Laor2008, Wang2024}, adopting the threshold $L_{\rm R}/L_{\rm X} < 10^{-2.73}$ for non-jetted systems. The final sample consists of 125 non-jetted AGNs with reliable variability measurements (see Table \ref{table1}).

\begin{table*}
	\caption{The sample of jetted AGNs.}
	\centering
	\label{table00}
	\setlength{\tabcolsep}{2.3mm}{
		\begin{tabular}{lllllllllllllllllllllllllllllll}% four columns, alignment for each
			\hline
			\hline
			Name &   RA &     Decl   &$z$   &   $\log \tau_{\rm DRW}$ & $\log \tau_{\rm DRW, le}$  & $\log \tau_{\rm DRW, ue}$  & $\log M_{\rm BH}$ & $\log M_{\rm BH, er}$ & $\log L_{\rm disk}$&  $\delta$ \\
			(1)    &  (2)  & (3)   & (4) & (5) & (6) & (7) & (8) & (9) & (10) & (11)\\
			\hline
			4FGL J0003.2+2207	&	0.8058	&	22.1302	&	0.1	&	1.027	&	-0.246	&	0.273	&	8.1	&	0.11	&	42.74	&	2.47	\\
			4FGL J0004.3+4614	&	1.0757	&	46.2427	&	1.81	&	1.42	&	-0.104	&	0.123	&	8.36	&	0.1	&	46.07	&	7.75	\\
			4FGL J0006.4+0135	&	1.6094	&	1.5982	&	0.78	&	1.622	&	-0.195	&	0.233	&	9.33	&	0.33	&	44.62	&	4.75	\\
			4FGL J0011.4+0057	&	2.8551	&	0.9646	&	1.49	&	1.337	&	-0.102	&	0.12	&	8.66	&	0.05	&	45.71	&	11.03	\\
			4FGL J0013.6-0424	&	3.4208	&	-4.4017	&	1.07	&	1.524	&	-0.151	&	0.18	&	7.82	&	0.08	&	45.03	&	27.16	\\
			4FGL J0017.8+1455	&	4.4711	&	14.9228	&	0.3	&	1.64	&	-0.129	&	0.162	&	8.27	&	0.44	&	44.16	&	3.79	\\
			4FGL J0021.6-0855	&	5.4115	&	-8.9174	&	0.64	&	2.081	&	-0.218	&	0.36	&	8.54	&	0.45	&	44.63	&	4.55	\\
			4FGL J0022.0+0006	&	5.5154	&	0.1134	&	0.3	&	1.894	&	-0.18	&	0.251	&	8.02	&	0.4	&	43.79	&	4.82	\\
			4FGL J0030.6-0212	&	7.6558	&	-2.2014	&	1.8	&	1.078	&	-0.091	&	0.102	&	8.66	&	0.28	&	45.98	&	7.44	\\
			4FGL J0033.3-2040	&	8.3457	&	-20.6674	&	0.07	&	1.396	&	-0.199	&	0.243	&	8.56	&	0.11	&	42.93	&	2.69	\\
			4FGL J0036.9+1832	&	9.234	&	18.5424	&	1.59	&	1.963	&	-0.241	&	0.317	&	8	&	0.09	&	45.7	&	5.89	\\
			4FGL J0038.2-2459	&	9.5589	&	-24.9941	&	0.49	&	1.744	&	-0.16	&	0.219	&	8.14	&	0.23	&	44.97	&	5.64	\\
			4FGL J0040.4-2340	&	10.1012	&	-23.6704	&	0.21	&	1.57	&	-0.229	&	0.253	&	8.68	&	0.13	&	43.75	&	3.31	\\
			4FGL J0059.3-0152	&	14.8361	&	-1.8725	&	0.14	&	1.63	&	-0.143	&	0.187	&	8.63	&	0.13	&	43.52	&	3.79	\\
			4FGL J0102.4+4214	&	15.6069	&	42.2371	&	0.87	&	1.043	&	-0.181	&	0.244	&	8.33	&	0.24	&	45.66	&	24.26	\\
			4FGL J0103.8+1321	&	15.969	&	13.3536	&	0.49	&	1.988	&	-0.179	&	0.258	&	9.69	&	0.25	&	44.41	&	4.52	\\
			4FGL J0109.3+2401	&	17.3313	&	24.0237	&	0.49	&	1.185	&	-0.12	&	0.131	&	8.22	&	0.45	&	44.28	&	4.44	\\
			4FGL J0151.3+8601	&	27.8381	&	86.0194	&	0.15	&	1.503	&	-0.179	&	0.217	&	9.25	&	0.22	&	43.54	&	3.32	\\
			4FGL J0152.2+2206	&	28.0727	&	22.1133	&	1.32	&	1.723	&	-0.133	&	0.172	&	8.45	&	0.06	&	46.17	&	4.32	\\
			4FGL J0222.0-1616	&	35.5197	&	-16.2787	&	0.69	&	1.169	&	-0.1	&	0.117	&	7.58	&	0.48	&	45.34	&	14.95	\\
			4FGL J0223.5-0928	&	35.8995	&	-9.481	&	1	&	1.473	&	-0.151	&	0.177	&	8.37	&	0.24	&	45.14	&	5	\\
			4FGL J0224.2+0700	&	36.0577	&	7.0126	&	0.51	&	1.634	&	-0.169	&	0.221	&	8.87	&	0.32	&	44.42	&	9.16	\\
			4FGL J0226.3-1845	&	36.5838	&	-18.763	&	1.67	&	1.785	&	-0.239	&	0.3	&	7.94	&	0.27	&	45.78	&	5.32	\\
			4FGL J0227.3+0201	&	36.8296	&	2.0203	&	0.45	&	1.903	&	-0.158	&	0.229	&	9.47	&	0.19	&	44.52	&	5.48	\\
			4FGL J0232.8+2018	&	38.2139	&	20.3159	&	0.13	&	1.88	&	-0.182	&	0.254	&	10.08	&	0.22	&	43.18	&	3.86	\\
			4FGL J0237.7+0206	&	39.4414	&	2.1019	&	0.022	&	1.273	&	-0.173	&	0.191	&	8.72	&	0.26	&	44.78	&		\\
			4FGL J0250.6+1712	&	42.6563	&	17.2081	&	0.24	&	2.154	&	-0.181	&	0.289	&	9.47	&	0.18	&	44.23	&	4.38	\\
			4FGL J0257.9-1215	&	44.492	&	-12.2637	&	1.39	&	1.782	&	-0.222	&	0.296	&	9.15	&	0.11	&	46.1	&	5.35	\\
			4FGL J0305.1-1608	&	46.2919	&	-16.1466	&	0.31	&	2.189	&	-0.265	&	0.428	&	9.24	&	0.15	&	44.1	&	4.67	\\
			4FGL J0309.0+1029	&	47.2617	&	10.4927	&	0.86	&	1.302	&	-0.161	&	0.18	&	7.77	&	0.16	&	44.76	&	13.74	\\
			4FGL J0310.9+3815	&	47.7297	&	38.2552	&	0.81	&	1.578	&	-0.137	&	0.206	&	8.19	&	0.3	&	45.09	&	34.97	\\
			4FGL J0312.8+0134	&	48.2222	&	1.5724	&	0.66	&	1.765	&	-0.215	&	0.277	&	8.17	&	0.2	&	45.3	&	11.77	\\
			4FGL J0315.9-1033	&	48.9922	&	-10.5522	&	1.56	&	1.175	&	-0.204	&	0.224	&	8.75	&	0.3	&	46.38	&	17.21	\\
			4FGL J0324.8+3412	&	51.2058	&	34.2119	&	0.06	&	2.006	&	-0.165	&	0.244	&	7.21	&	0.36	&	43.95	&	5.7	\\
			4FGL J0325.3+3332	&	51.3388	&	33.5383	&	0.12	&	1.282	&	-0.94	&	0.548	&	9.2	&	0.16	&	43.1	&	2.85	\\
			4FGL J0327.5-1805	&	51.8907	&	-18.0885	&	0.73	&	1.961	&	-0.271	&	0.365	&	8.04	&	0.22	&	44.96	&	4.07	\\
			4FGL J0401.7+2112	&	60.4496	&	21.2007	&	0.83	&	1.13	&	-0.09	&	0.103	&	8.43	&	0.36	&	45.15	&	25.76	\\
			4FGL J0423.3-0120	&	65.826	&	-1.3342	&	0.91	&	1.596	&	-0.11	&	0.138	&	8.4	&	0.23	&	45.65	&	5.31	\\
			4FGL J0501.2-0158	&	75.3023	&	-1.9749	&	2.29	&	1.663	&	-0.134	&	0.175	&	8.66	&	0	&	46.32	&	24.92	\\
			4FGL J0502.4+0609	&	75.6181	&	6.1627	&	1.11	&	2.127	&	-0.254	&	0.405	&	8.39	&	0.36	&	45.63	&	8.23	\\
			4FGL J0510.0+1800	&	77.5181	&	18.0135	&	0.41	&	1.774	&	-0.141	&	0.179	&	8.38	&	0.49	&	45.98	&	21.74	\\
			4FGL J0602.0+5315	&	90.5148	&	53.2657	&	0.05	&	1.762	&	-0.152	&	0.185	&	9.55	&	0.18	&	42.88	&	3.05	\\
			4FGL J0608.1-1521	&	92.0302	&	-15.3639	&	1.09	&	1.378	&	-0.275	&	0.286	&	8.17	&	0.32	&	45.86	&	29.5	\\
			4FGL J0629.3-1959	&	97.3478	&	-19.9999	&	1.71	&	1.709	&	-0.307	&	0.602	&	9.88	&	0.55	&	45.83	&	15.8	\\
			4FGL J0634.9-2335	&	98.7305	&	-23.5949	&	1.53	&	1.649	&	-0.245	&	0.375	&	8.37	&	0.28	&	45.35	&	5.81	\\
			4FGL J0739.2+0137	&	114.82	&	1.6216	&	0.19	&	1.284	&	-0.092	&	0.107	&	8.04	&	0.29	&	44.96	&	11.44	\\
			4FGL J0742.1+4902	&	115.5361	&	49.0428	&	2.31	&	2.098	&	-0.184	&	0.3	&	8.18	&	0.48	&	44.88	&	20.51	\\
			4FGL J0746.4+2546	&	116.6021	&	25.7678	&	2.99	&	1.948	&	-0.21	&	0.278	&	9.37	&	0.07	&	46.61	&	31.47	\\
			4FGL J0749.9+1823	&	117.4853	&	18.387	&	1.16	&	2.145	&	-0.232	&	0.348	&	8.41	&	0.07	&	45.7	&	4.92	\\
			\hline
		\end{tabular}
	}
	\footnotesize{
		Col. 1: name;
		Col. 2: the R.A. in decimal degrees; 
		Col. 3: (delineation) in decimal degrees;
		Col. 4: redshift;
		Col. 5-Col.7: the optical variability damping timescale and its corresponding lower and upper error in units days;
		Col.8-Col.9: the black hole mass and error;
		Col.10: the disk luminosity in units erg s$^{-1}$;
		Col.11: the Doppler factor. This table is available in its entirety in machine-readable form in the online article.} 
	\label{para00}
\end{table*}

\begin{table}
	\caption{The sample of Microquasar.}
	\centering
	\label{table0}
	\setlength{\tabcolsep}{3.0mm}{
		\begin{tabular}{lllllllllllllllllllllllllllllll}% four columns, alignment for each
			\hline
			\hline
			Name &   $\rm ln \tau_{\rm DRW}$  & $\log M_{\rm BH}$ & $\log M_{\rm BH, er}$ \\
			(1)    &  (2)  & (3)   & (4) \\
			\hline
			SS 433	&	$-4.91_{-0.22}^{+0.24}$	&	0.62	&	0.04	\\
			GRO J1655-40	&	$-4.14_{-0.37}^{+0.46}$	&	0.73	&	0.01	\\
			IGR J17091-3624	&	$-5.91_{-0.17}^{+0.17}$	&	1.1	&	0.09	\\
			LS I+61 303	&	$-4.58_{-0.63}^{+0.88}$	&	0.4	&	0.2	\\
			GRS 1915+105	&	$-3.33_{-0.48}^{+0.7}$	&	1.11	&	0.07	\\
			H1743-322	&	$-2.75_{-0.29}^{+0.39}$	&	1.04	&	0.07	\\
			Cygnus X-1	&	$-3.41_{-0.22}^{+0.27}$	&	1.17	&	0.02	\\
			\hline
		\end{tabular}
	}
	\footnotesize{
		Col. 1: name;
		Col. 2: the optical variability damping timescale and its corresponding lower and upper error in units days;
		Col.3-Col.4: the black hole mass and error.} 
	\label{para0}
\end{table}

\begin{table*}
	\caption{The sample of non-jetted AGNs.}
	\centering
	\label{table1}
	\setlength{\tabcolsep}{3.0mm}{
		\begin{tabular}{lllllllllllllllllllllllllllllll}% four columns, alignment for each
			\hline
			\hline
			Name &   RA &     Decl   &$z$   &  $S_{1.4}$ & $S_{\rm 1.4, error}$ & $S_{X}$ & $S_{Xer}$ & $\log \tau_{\rm DRW}$  & $\log M_{\rm BH}$ & $\log L_{\rm disk}$\\
			(1)    &  (2)  & (3)   & (4) & (5) & (6) & (7) & (8) & (9) & (10) & (11)\\
			\hline
J0002.5+0323	&	0.60941	&	3.35214	&	0.02545663	&	5.2	&	0.5	&	1.23E-11	&	2.23E-13	&	2.13$_{-0.18}^{+0.27}$	&	6.7	&	42.3	\\
J0034.6-0422	&	8.63667	&	-4.40347	&	0.2129818	&	1.03	&	0.5	&	1.78E-12	&	2.20E-13	&	1.85$_{-0.23}^{+0.26}$	&	9.22	&	43.9	\\
J0036.3+4540	&	9.08653	&	45.66476	&	0.04760781	&	336.15	&	39.63	&	1.31E-11	&	2.79E-13	&	2.02$_{-0.14}^{+0.19}$	&	7.36	&	43.04	\\
J0041.0+2444	&	10.16581	&	24.76064	&	0.07836469	&	5.9	&	1.4	&	2.72E-12	&	1.76E-13	&	1.59$_{-0.23}^{+0.26}$	&	7.73	&	43.06	\\
J0051.6+2928	&	12.89537	&	29.4008	&	0.03589227	&	10	&	0.5	&	6.28E-12	&	1.93E-13	&	2.23$_{-0.2}^{+0.3}$	&	7.62	&	42.5	\\
J0051.9+1724	&	12.97793	&	17.43323	&	0.06402702	&	5.526	&	0.336	&	2.63E-11	&	2.20E-13	&	2.34$_{-0.22}^{+0.37}$	&	7.75	&	43.76	\\
J0059.4+3150	&	14.97188	&	31.82691	&	0.01463096	&	1.85	&	0.263	&	1.38E-11	&	1.50E-13	&	1.68$_{-0.09}^{+0.12}$	&	7.56	&	41.77	\\
J0105.7-1414	&	16.41112	&	-14.271	&	0.06676629	&	2.5	&	0	&	1.28E-11	&	1.78E-13	&	2.36$_{-0.2}^{+0.33}$	&	7.88	&	43.55	\\
J0113.8+1313	&	18.46233	&	13.27181	&	0.049926	&	11.7	&	0.6	&	3.63E-12	&	2.32E-13	&	1.98$_{-0.15}^{+0.2}$	&	7.45	&	42.85	\\
J0116.5-1235	&	19.1297	&	-12.60485	&	0.14244739	&	3.9	&	0.5	&	6.13E-12	&	2.41E-13	&	1.91$_{-0.22}^{+0.25}$	&	7.56	&	43.28	\\
J0122.8+5003	&	20.6428	&	50.05449	&	0.02058234	&	10.2	&	0.5	&	1.42E-12	&	1.02E-13	&	1.51$_{-0.2}^{+0.23}$	&	7.54	&	43.28	\\
J0138.8+2925	&	24.84983	&	29.40158	&	0.07180081	&	4.6	&	0.4	&	3.37E-12	&	1.48E-13	&	1.95$_{-0.21}^{+0.26}$	&	7.78	&	42.99	\\
J0149.2+2153B	&	27.35337	&	21.99727	&	0.00948535	&	5.3	&	0.5	&	8.14E-13	&	1.04E-13	&	1.07$_{-0.31}^{+0.28}$	&	8.28	&	41.43	\\
J0157.2+4715	&	29.29571	&	47.26673	&	0.04827464	&	2.5	&	0	&	1.46E-11	&	2.75E-13	&	2.16$_{-0.17}^{+0.24}$	&	7.88	&	42.39	\\
J0206.2-0019	&	31.56665	&	-0.29138	&	0.04263127	&	4.2	&	0.4	&	1.42E-11	&	1.33E-13	&	2.04$_{-0.22}^{+0.28}$	&	7.81	&	42.78	\\
J0208.5-1738	&	32.14546	&	-17.6596	&	0.12909222	&	39.4	&	1.8	&	9.14E-12	&	2.40E-13	&	2.34$_{-0.21}^{+0.38}$	&	9.02	&	43.92	\\
J0228.1+3118	&	37.06014	&	31.31119	&	0.01630788	&	13.3	&	1.1	&	3.78E-11	&	6.85E-13	&	1.12$_{-0.19}^{+0.24}$	&	7.41	&	41.73	\\
J0230.2-0900	&	37.52302	&	-8.99808	&	0.01727102	&	2.4	&	0.5	&	2.50E-11	&	1.85E-13	&	1.67$_{-0.12}^{+0.16}$	&	6.45	&	42.41	\\
J0234.6-0848	&	38.65758	&	-8.78813	&	0.04302233	&	15.3	&	1	&	1.90E-11	&	1.72E-13	&	1.53$_{-0.15}^{+0.19}$	&	8.11	&	43.36	\\
J0241.6+0711	&	40.39481	&	7.18686	&	0.02730802	&	1.48	&	0.41	&	5.97E-12	&	1.63E-13	&	1.95$_{-0.18}^{+0.23}$	&	7.41	&	42.09	\\
J0251.3+5441	&	42.67706	&	54.7049	&	0.01527144	&	8.5	&	0.5	&	8.51E-13	&	1.11E-13	&	1.67$_{-0.33}^{+0.34}$	&	7.44	&	42.03	\\
J0324.9+4044	&	51.30307	&	40.69865	&	0.04724702	&	8.1	&	0.5	&	5.70E-12	&	2.88E-13	&	1.33$_{-0.23}^{+0.25}$	&	6.64	&	42.35	\\
J0414.8-0754	&	63.71977	&	-7.92758	&	0.03824158	&	8.9	&	0.5	&	7.33E-12	&	3.11E-13	&	1.64$_{-0.25}^{+0.31}$	&	8.05	&	43.81	\\
J0429.6-2114	&	67.40967	&	-21.16232	&	0.07000987	&	10.3	&	0.6	&	9.08E-12	&	2.09E-13	&	1.8$_{-0.22}^{+0.26}$	&	8.82	&	43.66	\\
J0440.9+2741	&	70.19884	&	27.66323	&	0.03638922	&	3.1	&	0.6	&	5.33E-12	&	8.79E-14	&	1.63$_{-0.25}^{+0.28}$	&	8.42	&	43.02	\\
J0441.2-2704	&	70.34399	&	-27.13862	&	0.08414478	&	30	&	1	&	1.60E-11	&	1.87E-13	&	1.84$_{-0.2}^{+0.24}$	&	8.04	&	44.19	\\
J0441.8-0823	&	70.47525	&	-8.44281	&	0.04327938	&	1.82	&	0.54	&	8.45E-12	&	1.84E-13	&	2.02$_{-0.22}^{+0.31}$	&	6.78	&	42.95	\\
J0459.7+3502	&	74.98664	&	35.04797	&	0.0443829	&	10.2	&	0.5	&	4.52E-12	&	2.17E-13	&	1.35$_{-0.33}^{+0.33}$	&	7.39	&	42.85	\\
J0502.1+0332	&	75.53714	&	3.5303	&	0.01588625	&	18.7	&	0.7	&	9.04E-12	&	1.73E-13	&	1.85$_{-0.13}^{+0.16}$	&	7.02	&	42.76	\\
J0503.0+2302	&	75.742	&	22.99732	&	0.05816045	&	23.6	&	0.8	&	2.07E-11	&	3.26E-13	&	2.3$_{-0.19}^{+0.31}$	&	8.35	&	43.98	\\
J0510.7+1629	&	77.69	&	16.499	&	0.01735243	&	6	&	0.5	&	2.31E-11	&	4.00E-14	&	2.28$_{-0.17}^{+0.29}$	&	6.88	&	42.96	\\
J0516.2-0009	&	79.04725	&	-0.14951	&	0.03256885	&	12.3	&	0.6	&	5.77E-11	&	4.71E-13	&	2.19$_{-0.22}^{+0.37}$	&	8.07	&	43.46	\\
J0516.4-1034	&	79.08867	&	-10.56208	&	0.02805341	&	3.4	&	0	&	8.85E-12	&	1.51E-13	&	2.2$_{-0.19}^{+0.27}$	&	7.02	&	43.07	\\
J0524.1-1210	&	81.02887	&	-12.16916	&	0.04918616	&	3	&	0	&	3.50E-12	&	1.95E-13	&	2.07$_{-0.15}^{+0.21}$	&	8.27	&	43.67	\\
J0535.4+4013	&	83.88305	&	40.1876	&	0.01994829	&	8.5	&	0.5	&	4.58E-12	&	1.48E-13	&	1.52$_{-0.19}^{+0.26}$	&	6.59	&	42.87	\\
J0623.8+6445	&	95.89618	&	64.76011	&	0.08649497	&	3.5	&	0.5	&	4.72E-12	&	1.57E-13	&	2.26$_{-0.17}^{+0.27}$	&	7.28	&	43.43	\\
J0626.6+0729	&	96.61276	&	7.45833	&	0.04244714	&	2.74	&	0.7	&	2.87E-12	&	1.59E-13	&	2.24$_{-0.21}^{+0.31}$	&	7.8	&	42.71	\\
J0630.7+6342	&	98.1979	&	63.67341	&	0.01271338	&	11.6	&	0.5	&	1.28E-11	&	5.57E-13	&	2.32$_{-0.22}^{+0.37}$	&	6.06	&	42.44	\\
J0640.4-2554	&	100.04853	&	-25.89504	&	0.02524503	&	38.5	&	1.6	&	2.88E-11	&	3.34E-13	&	2.12$_{-0.29}^{+0.37}$	&	7.15	&	43.24	\\
J0655.8+3957	&	103.95544	&	40.0002	&	0.01719824	&	4.7	&	0.4	&	1.06E-11	&	4.23E-13	&	1.39$_{-0.09}^{+0.1}$	&	7.33	&	43.3	\\
J0727.4-2408	&	111.83775	&	-24.10904	&	0.12191061	&	29.1	&	1	&	7.07E-12	&	3.92E-13	&	1.99$_{-0.46}^{+0.51}$	&	7.67	&	44.51	\\
J0736.9+5846	&	114.23779	&	58.77022	&	0.03988829	&	3.6	&	0.5	&	1.13E-11	&	1.88E-13	&	1.52$_{-0.1}^{+0.11}$	&	7.6	&	43.14	\\
J0742.5+4948	&	115.6375	&	49.80939	&	0.0221083	&	20.5	&	1	&	2.71E-11	&	5.63E-13	&	1.94$_{-0.12}^{+0.16}$	&	7.61	&	42.55	\\
J0747.5+6057	&	116.86988	&	60.93349	&	0.02887968	&	5.8	&	0.5	&	1.28E-11	&	1.77E-13	&	1.88$_{-0.15}^{+0.18}$	&	7.19	&	42.31	\\
J0817.8+0251	&	124.45484	&	2.84998	&	0.1060687	&	336.15	&	39.63	&	8.64E-12	&	2.40E-13	&	2.2$_{-0.16}^{+0.25}$	&	7.83	&	44.15	\\
J0818.1+0120	&	124.56056	&	1.37418	&	0.08971186	&	20.4	&	1.1	&	9.50E-12	&	2.42E-13	&	2.3$_{-0.33}^{+0.5}$	&	7.67	&	43.3	\\
J0845.3+1420	&	131.32694	&	14.34307	&	0.0607264	&	7.4	&	0.5	&	7.78E-12	&	2.22E-13	&	2.05$_{-0.25}^{+0.33}$	&	8.64	&	43.51	\\
J0904.3+5538	&	136.15378	&	55.60017	&	0.03714255	&	2.2	&	0.4	&	8.30E-12	&	2.00E-13	&	1.77$_{-0.12}^{+0.15}$	&	7.72	&	42.92	\\
J0918.5+1618	&	139.60833	&	16.30559	&	0.0294596	&	6.1	&	0.5	&	1.43E-11	&	2.09E-13	&	1.54$_{-0.1}^{+0.12}$	&	8.24	&	43.39	\\
			\hline
	\end{tabular}}
\end{table*}
\addtocounter{table}{-1}
\begin{table*}
	\caption{$Continue.$}
	\centering
	\setlength{\tabcolsep}{3.0mm}{
		\begin{tabular}{lllllllllllllllllllllllllllllll}% four columns, alignment for each
			\hline
			\hline
			Name &   RA &     Decl   &$z$   &  $S_{1.4}$ & $S_{\rm 1.4, error}$ & $S_{X}$ & $S_{Xer}$ & $\log \tau_{\rm DRW}$  & $\log M_{\rm BH}$ & $\log L_{\rm disk}$\\
			(1)    &  (2)  & (3)   & (4) & (5) & (6) & (7) & (8) & (9) & (10) & (11)\\
			\hline
J0926.2+1244	&	141.51364	&	12.73462	&	0.02862431	&	8.526	&	0	&	2.64E-11	&	5.97E-13	&	1.96$_{-0.16}^{+0.2}$	&	7.08	&	43.12	\\
J0927.3+2301	&	141.82655	&	23.02026	&	0.02643901	&	2.792	&	0	&	7.32E-12	&	2.19E-13	&	1.83$_{-0.2}^{+0.23}$	&	7.91	&	42.37	\\
J0951.9-0649	&	147.97887	&	-6.82244	&	0.01441361	&	1.476	&	0	&	1.17E-11	&	2.47E-13	&	2.07$_{-0.28}^{+0.36}$	&	7.24	&	42.06	\\
J1023.5+1952	&	155.8775	&	19.86457	&	0.00327673	&	97.5	&	3.5	&	6.92E-11	&	2.72E-13	&	1.7$_{-0.14}^{+0.16}$	&	6.77	&	41.71	\\
J1031.9-1418	&	157.97627	&	-14.28119	&	0.08519689	&	14.3	&	0.7	&	4.34E-11	&	5.13E-13	&	1.82$_{-0.19}^{+0.35}$	&	8.85	&	43.74	\\
J1043.4+1105	&	160.86054	&	11.08962	&	0.04766739	&	336.15	&	39.63	&	1.05E-11	&	2.20E-13	&	2.02$_{-0.14}^{+0.2}$	&	7.89	&	43.22	\\
J1100.9+1104	&	165.25684	&	11.04722	&	0.03574203	&	2.859	&	0	&	1.02E-11	&	3.42E-13	&	2.08$_{-0.21}^{+0.28}$	&	6.66	&	42.22	\\
J1103.4+3731	&	165.91755	&	37.49039	&	0.07387791	&	5.3	&	0.4	&	6.46E-12	&	1.99E-13	&	2.09$_{-0.2}^{+0.27}$	&	7.67	&	42.82	\\
J1106.5+7234	&	166.6966	&	72.56857	&	0.00871832	&	31.3	&	1.3	&	3.41E-11	&	2.12E-13	&	2.25$_{-0.24}^{+0.34}$	&	7.39	&	42.13	\\
J1125.6+5423	&	171.40027	&	54.38179	&	0.02061314	&	1.533	&	0	&	1.83E-11	&	2.89E-13	&	1.73$_{-0.12}^{+0.13}$	&	6.67	&	42.35	\\
J1132.9+1019A	&	173.20471	&	10.29647	&	0.0437172	&	336.15	&	39.63	&	7.52E-12	&	1.87E-13	&	1.32$_{-0.18}^{+0.19}$	&	7.7	&	42.55	\\
J1136.0+2132	&	174.12234	&	21.59594	&	0.02946955	&	10	&	0.5	&	1.45E-11	&	2.57E-13	&	1.86$_{-0.15}^{+0.18}$	&	6.99	&	42.76	\\
J1139.0-2323	&	174.7127	&	-23.35931	&	0.02717489	&	8.2	&	0.5	&	1.31E-11	&	1.20E-13	&	1.91$_{-0.2}^{+0.25}$	&	7.84	&	42.67	\\
J1139.1+5913	&	174.78736	&	59.19856	&	0.06121367	&	1.65	&	0.4	&	1.74E-11	&	2.09E-13	&	2.2$_{-0.17}^{+0.26}$	&	7.85	&	43.18	\\
J1141.3+2156	&	175.31719	&	21.93944	&	0.06320943	&	4.5	&	0.5	&	1.37E-11	&	1.90E-13	&	2.08$_{-0.29}^{+0.37}$	&	7.39	&	43.78	\\
J1142.2+1021	&	175.54612	&	10.27729	&	0.01947938	&	34	&	1.1	&	6.68E-12	&	2.13E-13	&	1.83$_{-0.21}^{+0.24}$	&	7.43	&	41.76	\\
J1144.1+3652	&	176.12393	&	36.88596	&	0.03828111	&	1.14	&	0.25	&	5.87E-12	&	9.55E-14	&	2.07$_{-0.13}^{+0.2}$	&	8.11	&	42.46	\\
J1148.3+0901	&	176.97911	&	9.04146	&	0.06894126	&	1.147	&		&	7.44E-12	&	2.02E-13	&	2.3$_{-0.18}^{+0.3}$	&	8.24	&	43.44	\\
J1201.2-0341	&	180.30944	&	-3.67768	&	0.01946029	&	3.4	&	0.5	&	1.21E-11	&	1.81E-13	&	1.71$_{-0.19}^{+0.22}$	&	6.21	&	42.5	\\
J1203.0+4433	&	180.79001	&	44.53131	&	0.0020431	&	94.4	&	4	&	3.44E-11	&	1.80E-13	&	1.08$_{-0.07}^{+0.08}$	&	6.13	&	41.26	\\
J1205.8+4959	&	181.48284	&	49.99855	&	0.06311923	&	1.794	&		&	2.64E-12	&	9.67E-14	&	1.36$_{-0.1}^{+0.11}$	&	8.32	&	43.24	\\
J1218.5+2952	&	184.610475	&	29.81294	&	0.01292423	&	38.1	&	1.2	&	1.70E-11	&	2.00E-14	&	1.25$_{-0.09}^{+0.1}$	&	6.82	&	43	\\
J1219.4+4720	&	184.7399	&	47.30356	&	0.00169195	&	5.926	&		&	3.58E-11	&	1.03E-12	&	1.59$_{-0.43}^{+0.42}$	&	7.56	&	40.47	\\
J1223.7+0238	&	185.85003	&	2.6793	&	0.02368976	&	336.15	&	39.63	&	2.07E-11	&	1.95E-13	&	2.21$_{-0.18}^{+0.28}$	&	7.42	&	41.97	\\
J1302.9+1620	&	195.74494	&	16.40819	&	0.06700358	&	32.9	&	1.1	&	1.18E-11	&	2.70E-13	&	1.79$_{-0.12}^{+0.16}$	&	7.3	&	44.11	\\
J1309.2+1139	&	197.27308	&	11.63398	&	0.02518155	&	1.64	&	0.15	&	2.53E-12	&	3.27E-13	&	1.86$_{-0.21}^{+0.24}$	&	8.59	&	41.52	\\
J1313.1-1108	&	198.27366	&	-11.12832	&	0.03458327	&	2.5	&	0.6	&	1.75E-11	&	2.52E-13	&	1.63$_{-0.14}^{+0.17}$	&	7.49	&	43.14	\\
J1313.6+3650A	&	198.45418	&	36.89912	&	0.06694541	&	1.273	&	0.193	&	3.24E-12	&	2.45E-13	&	1.31$_{-0.14}^{+0.16}$	&	7.74	&	43.08	\\
J1331.2-2524	&	202.80764	&	-25.40219	&	0.02607113	&	12.6	&	0.6	&	5.26E-12	&	1.76E-13	&	1.96$_{-0.19}^{+0.24}$	&	8.32	&	42.94	\\
J1341.2-1439	&	205.30328	&	-14.64449	&	0.04175325	&	5.3	&	0.6	&	2.51E-11	&	3.69E-13	&	2.1$_{-0.15}^{+0.22}$	&	8.03	&	43.5	\\
J1341.9+3537	&	205.53563	&	35.65439	&	0.00360589	&	3.5	&	0.4	&	2.49E-11	&	7.42E-13	&	2.09$_{-0.15}^{+0.21}$	&	6.66	&	40.84	\\
J1349.7+0209	&	207.47001	&	2.07913	&	0.03275976	&	0.848	&	0	&	1.00E-11	&	2.62E-13	&	1.82$_{-0.15}^{+0.19}$	&	7.16	&	42.9	\\
J1351.5-1814	&	207.87249	&	-18.22988	&	0.01208548	&	336.15	&	39.63	&	2.17E-11	&	2.75E-13	&	1.92$_{-0.22}^{+0.26}$	&	6.11	&	42.45	\\
J1356.6-1932	&	209.15281	&	-19.52913	&	0.03497423	&	12.6	&	0.6	&	1.66E-11	&	3.88E-13	&	1.86$_{-0.15}^{+0.19}$	&	7.61	&	42.95	\\
J1417.9+2507	&	214.49824	&	25.13668	&	0.01668582	&	28.2	&	1.2	&	4.82E-11	&	9.49E-14	&	2.11$_{-0.15}^{+0.23}$	&	7.72	&	42.82	\\
J1419.0-2639	&	214.843352	&	-26.644752	&	0.02294033	&	12.5	&	1.2	&	1.37E-11	&	2.00E-14	&	2.25$_{-0.21}^{+0.34}$	&	7.23	&	41.78	\\
J1424.2+2435	&	216.09548	&	24.61407	&	0.01691813	&	31.4	&	1.3	&	4.67E-12	&	1.70E-13	&	1.95$_{-0.45}^{+0.35}$	&	7.71	&	42.36	\\
J1431.2+2816	&	217.76974	&	28.28718	&	0.04600112	&	3.2	&	0.4	&	7.02E-12	&	9.65E-14	&	2.22$_{-0.17}^{+0.26}$	&	6.78	&	42.69	\\
J1434.9+4837	&	218.71996	&	48.66207	&	0.0364747	&	2	&	0.1	&	1.11E-12	&	8.76E-14	&	1.93$_{-0.26}^{+0.3}$	&	7.69	&	42.56	\\
J1436.4+5846	&	219.09205	&	58.79413	&	0.03119806	&	11.2	&	0.5	&	2.06E-11	&	7.34E-14	&	2.27$_{-0.16}^{+0.25}$	&	7.59	&	41.91	\\
J1453.3+2558	&	223.28297	&	25.90923	&	0.04858042	&	0.94	&	0.32	&	1.98E-11	&	3.41E-13	&	2.3$_{-0.19}^{+0.32}$	&	8.48	&	42.64	\\
J1504.2+1025	&	226.00502	&	10.43778	&	0.03660376	&	3.44	&	0.56	&	2.65E-11	&	9.30E-14	&	2.18$_{-0.16}^{+0.23}$	&	8.16	&	43.37	\\
J1508.8-0013	&	227.22484	&	-0.19707	&	0.05438563	&	22.9	&	0.8	&	8.84E-12	&	1.25E-13	&	2.23$_{-0.21}^{+0.33}$	&	8.4	&	43.41	\\
J1512.0-2119	&	227.99852	&	-21.3171	&	0.04446332	&	46.9	&	1.5	&	1.87E-11	&	6.01E-13	&	2.35$_{-0.23}^{+0.37}$	&	6.94	&	43.78	\\
J1521.8+0334	&	230.41555	&	3.62488	&	0.12629508	&	2.52	&	0.7	&	4.41E-12	&	1.29E-13	&	2.02$_{-0.15}^{+0.19}$	&	8.52	&	43.81	\\
J1542.0-1410	&	235.59966	&	-14.18499	&	0.09671079	&	1.93	&	0.35	&	5.46E-12	&	2.38E-13	&	2.14$_{-0.26}^{+0.32}$	&	8.73	&	43.17	\\
J1553.6+2347	&	238.43201	&	23.80731	&	0.11754874	&	66.11	&	0	&	4.50E-12	&	1.70E-13	&	2.12$_{-0.19}^{+0.28}$	&	9.2	&	42.77	\\
J1607.2+4834	&	241.80807	&	48.55711	&	0.12485957	&	1.71	&	0.15	&	6.85E-12	&	2.11E-13	&	2.14$_{-0.22}^{+0.3}$	&	8.52	&	42.99	\\
J1621.2+8104	&	244.82929	&	81.04652	&	0.02392409	&	3.7	&	0.5	&	3.48E-12	&	1.79E-13	&	1.91$_{-0.2}^{+0.27}$	&	8.34	&	41.95	\\
J1708.6+2155	&	257.2466	&	21.88555	&	0.0725173	&	1	&	0	&	6.09E-12	&	2.00E-13	&	1.93$_{-0.12}^{+0.16}$	&	8.23	&	42.73	\\
			\hline
	\end{tabular}}
\end{table*}
\addtocounter{table}{-1}
\begin{table*}
	\caption{$Continue.$}
	\centering
	\setlength{\tabcolsep}{3.0mm}{
		\begin{tabular}{lllllllllllllllllllllllllllllll}% four columns, alignment for each
			\hline
			\hline
			Name &   RA &     Decl   &$z$   &  $S_{1.4}$ & $S_{\rm 1.4, error}$ & $S_{X}$ & $S_{Xer}$ & $\log \tau_{\rm DRW}$  & $\log M_{\rm BH}$ & $\log L_{\rm disk}$\\
			(1)    &  (2)  & (3)   & (4) & (5) & (6) & (7) & (8) & (9) & (10) & (11) \\
			\hline
J1719.7+4900	&	259.81059	&	48.98061	&	0.02486481	&	145.1	&	4.4	&	5.72E-12	&	1.65E-13	&	1.09$_{-0.09}^{+0.1}$	&	8	&	42.22	\\
J1723.5+3630	&	260.84663	&	36.5025	&	0.03998859	&	4.4	&	0.4	&	1.50E-11	&	3.72E-13	&	1.53$_{-0.1}^{+0.12}$	&	7.24	&	43.3	\\
J1733.3+3635	&	263.40366	&	36.5249	&	0.04366094	&	4.4	&	0.4	&	2.39E-12	&	1.60E-13	&	1.15$_{-0.12}^{+0.14}$	&	8.39	&	42.43	\\
J1741.9-1211	&	265.48024	&	-12.19937	&	0.03759909	&	3.5	&	0.5	&	1.72E-11	&	4.21E-13	&	2.09$_{-0.15}^{+0.24}$	&	8.03	&	43.02	\\
J1747.8+6837B	&	266.74782	&	68.60817	&	0.0637853	&	336.15	&	39.63	&	6.64E-12	&	1.82E-13	&	1.9$_{-0.12}^{+0.16}$	&	7.43	&	43.91	\\
J1824.2+1845	&	276.04504	&	18.76898	&	0.06669989	&	4.9	&	0.4	&	3.37E-12	&	1.59E-13	&	2.06$_{-0.18}^{+0.24}$	&	8.12	&	43.49	\\
J1830.8+0928	&	277.711	&	9.47842	&	0.01926842	&	3.65	&	0.98	&	1.39E-12	&	1.34E-13	&	2.31$_{-0.23}^{+0.35}$	&	7.43	&	41.21	\\
J1926.9+4140	&	291.62592	&	41.55137	&	0.07168564	&	2.18	&	0.28	&	8.75E-12	&	1.31E-13	&	2.27$_{-0.19}^{+0.31}$	&	8.09	&	43.01	\\
J1937.5-0613	&	294.38786	&	-6.2179	&	0.01035936	&	42.2	&	1.7	&	7.05E-11	&	5.28E-13	&	1.34$_{-0.13}^{+0.15}$	&	6.61	&	43.4	\\
J2015.2+2526	&	303.74696	&	25.38398	&	0.04589085	&	2.19	&	0.57	&	7.59E-13	&	9.80E-14	&	2.02$_{-0.18}^{+0.24}$	&	7.36	&	43.19	\\
J2035.2+2604	&	308.773	&	26.05798	&	0.0486436	&	9.5	&	0.5	&	7.97E-12	&	1.60E-13	&	2.09$_{-0.23}^{+0.32}$	&	7.18	&	43.48	\\
J2059.6+4301B	&	315.00395	&	43.03621	&	0.06602313	&	3.8	&	0.6	&	1.14E-12	&	1.28E-13	&	1.6$_{-0.18}^{+0.21}$	&	7.67	&	42.67	\\
J2109.1-0942	&	317.29137	&	-9.67087	&	0.02677063	&	5.3	&	0.4	&	2.24E-11	&	3.15E-13	&	1.92$_{-0.13}^{+0.17}$	&	7.51	&	-99	\\
J2116.3+2512	&	319.04297	&	25.28331	&	0.15342948	&	1.9	&	0.51	&	7.19E-12	&	2.41E-13	&	2.3$_{-0.31}^{+0.58}$	&	8.82	&	43.23	\\
J2138.8-3207	&	324.63886	&	32.0844	&	0.02468845	&	6.9	&	0.4	&	3.86E-12	&	1.70E-13	&	1.85$_{-0.12}^{+0.15}$	&	7.44	&	43.2	\\
J2145.5+1101	&	326.38665	&	11.04867	&	0.20864308	&	1.1025	&		&	4.12E-12	&	1.20E-13	&	2.13$_{-0.14}^{+0.21}$	&	8.17	&	43.74	\\
J2156.1+4728	&	323.97502	&	47.47283	&	0.02575057	&	6.8	&	0.5	&	1.19E-11	&	2.36E-13	&	1.89$_{-0.13}^{+0.16}$	&	7.4	&	42.83	\\
J2209.1-2747	&	332.28167	&	-27.80944	&	0.02266524	&	28.2	&	1.6	&	1.27E-11	&	1.59E-13	&	2.07$_{-0.17}^{+0.24}$	&	7.44	&	42.17	\\
J2214.2-2557	&	333.53808	&	-25.96379	&	0.05170644	&	3.1	&	0.5	&	1.66E-12	&	2.35E-13	&	1.46$_{-0.5}^{+0.46}$	&	7.96	&	42.5	\\
J2229.9+6646	&	337.3068	&	66.78074	&	0.11175124	&	381.4	&	14	&	9.12E-12	&	2.01E-13	&	2.25$_{-0.2}^{+0.3}$	&	8.91	&	43.45	\\
J2240.2+0801	&	340.07186	&	8.05417	&	0.02493767	&	15.1	&	0.6	&	7.34E-12	&	2.14E-13	&	2.23$_{-0.19}^{+0.29}$	&	7.08	&	42.88	\\
J2258.9+4054	&	344.73012	&	40.93229	&	0.01775995	&	11.8	&	0.9	&	1.68E-12	&	2.65E-13	&	1.18$_{-0.13}^{+0.13}$	&	8.6	&	41.7	\\
J2259.7+2458	&	344.88698	&	24.91821	&	0.03370706	&	3.7	&	0.4	&	1.99E-11	&	2.80E-13	&	1.55$_{-0.2}^{+0.24}$	&	6.87	&	43.59	\\
J2302.1+1557	&	345.504003	&	15.964772	&	0.00634269	&	20	&	1	&	3.26E-12	&	5.39E-14	&	1.66$_{-0.2}^{+0.23}$	&	7.02	&	41.3	\\
J2322.5-0646	&	350.60247	&	-6.76054	&	0.0329256	&	1.748	&	0	&	2.64E-13	&	6.11E-14	&	1.47$_{-0.28}^{+0.26}$	&	7.58	&	42.84	\\
J2352.6-1707	&	358.2141	&	-17.07705	&	0.0546952	&	1.3	&	0.33	&	1.14E-11	&	2.68E-13	&	1.91$_{-0.38}^{+0.55}$	&	8.29	&	43.09	\\
			\hline
		\end{tabular}
	}
	\footnotesize{
		Col. 1: name;
		Col. 2: the R.A. in decimal degrees; 
		Col. 3: (delineation) in decimal degrees;
		Col. 4: redshift;
		Col. 5: 1.4 GHz radio flux in units mjy; 
		Col. 6: 1.4 GHz radio flux error in units mjy;
		Col. 7: X-ray flux in units erg s$^{-1}$ cm$^{-2}$;
		Col. 8: X-ray flux error in units erg s$^{-1}$ cm$^{-2}$;
		Col. 9: the optical variability damping timescale and error in units days ;
		Col.10: the black hole mass;
		Col.11: the accretion disk luminosity in units erg s$^{-1}$. The accretion disk luminosity is from the work of Oh et al.(2022).
	} 
	\label{para}
\end{table*}

\section{Results and Discussion}
\begin{figure*}
	\includegraphics[width=16cm,height=18cm]{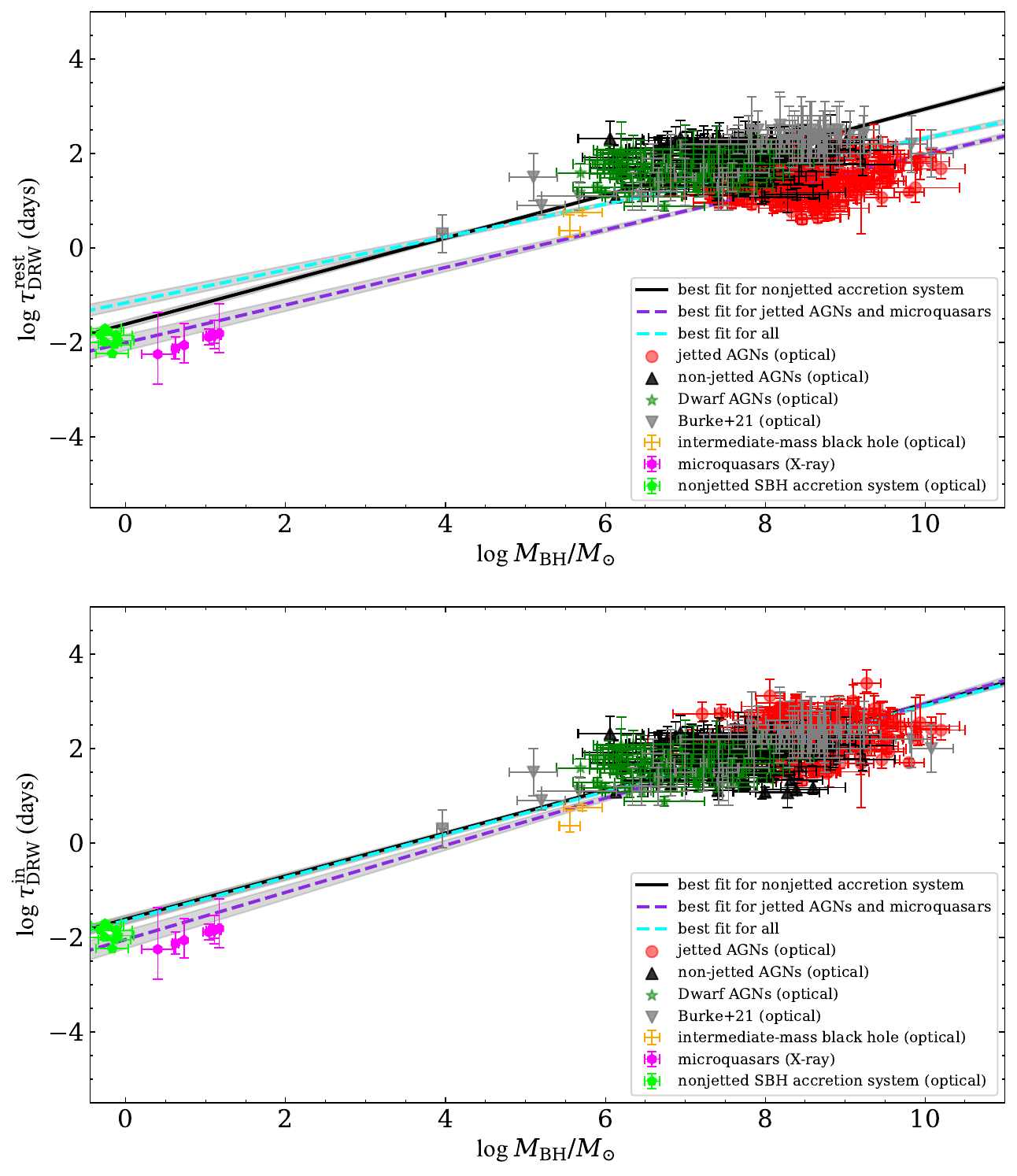}
	\centering
	\caption{Relation between the rest-frame optical variability timescale (top panel) and the intrinsic optical variability timescale (bottom panel) with black hole mass for our sample. The non-jetted stellar-mass black hole (SBH) accretion systems are from Burke et al.(2021). The microquasar sample is taken from Zhang et al.(2024). The dwarf active galactic nuclei (AGNs) are compiled from Wang et al.(2023). The intermediate-mass black hole (IMBH) data point is based on Su et al.(2024). The purple dashed line shows a linear fit to all jetted systems. The black dashed line represents the best fit for non-jetted systems. The cyan dashed line indicates the best fit for all sources combined. The shaded regions denote the 1$\sigma$ confidence intervals.}
	\label{tauMBH}
\end{figure*}

We perform a linear regression analysis between the optical variability timescale in the rest frame ($\tau_{\rm{DRW}}^{\rm rest}$) and black hole mass ($M_{\rm BH}$), accounting for measurement uncertainties in both variables. The relation is fitted using {\sffamily linmix}\footnote{https://linmix.readthedocs.io/en/latest/} \citep{Kelly2007}, a Bayesian method that properly handles observational errors and intrinsic scatter. The correlation between the rest-frame optical variability timescale and black hole mass for our sample is presented in the top panel of Figure~\ref{tauMBH}. A significant positive correlation is found across all sources, with a Pearson correlation coefficient of $r = 0.71$ and a highly significant $P$-value of $2.3 \times 10^{-73}$ (where $P < 0.05$ indicates statistical significance at the 95\% confidence level). The best-fitting result yields

\begin{equation}
	\log \tau_{\rm DRW}^{\rm rest} = (0.35\pm0.02)\log M_{\rm BH}+(-1.15\pm0.12),
\end{equation}        
with a $1\sigma$ intrinsic scatter of $0.22$ dex. We also investigate the relation between the rest-frame optical variability timescale and black hole mass for jetted and non-jetted systems separately. A significant positive correlation is found between the rest-frame optical variability timescale and black hole mass in jetted systems ($r = 0.85$, $P = 7.1 \times 10^{-55}$),

\begin{equation}
	\log \tau_{\rm DRW}^{\rm rest} = (0.40\pm0.02)\log M_{\rm BH}+(-2.01\pm0.17),
\end{equation}        
with a $1\sigma$ intrinsic scatter of $0.09$ dex. A significant positive correlation is also present between the rest-frame optical variability timescale and black hole mass for non-jetted systems ($r = 0.86$, $P = 1.4 \times 10^{-84}$),

\begin{equation}
	\log \tau_{\rm DRW}^{\rm rest} = (0.46\pm0.01)\log M_{\rm BH}+(-1.61\pm0.09),
\end{equation}         
with a $1\sigma$ intrinsic scatter of $0.06$ dex. From the top panel of Figure~\ref{tauMBH}, we observe that jetted AGNs tend to exhibit shorter rest-frame optical variability timescales compared to non-jetted AGNs. Due to the strong beaming effect in jetted AGNs, we further examine the relationship between the intrinsic optical variability timescale and black hole mass. However, owing to the lack of Doppler factor measurements and the unresolved location of the emission region, we do not correct for the beaming effect in the seven microquasars included in our sample. Moreover, several studies suggest that this effect may be negligible in microquasar systems \citep{Liodakis2017, Abeysekara2018}. The correlation between the intrinsic optical variability timescale and black hole mass for our sample is displayed in the bottom panel of Figure~\ref{tauMBH}. A significant positive correlation is found across all sources ($r = 0.86$, $P = 1.7 \times 10^{-142}$). The best-fitting result yields

\begin{equation}
	\log \tau_{\rm DRW}^{\rm in} = (0.45\pm0.01)\log M_{\rm BH}+(-1.62\pm0.09),
\label{tauinall}
\end{equation}        
with a $1\sigma$ intrinsic scatter of $0.11$ dex. A significant positive correlation is also present between the intrinsic optical variability timescale and black hole mass for jetted AGNs ($r = 0.85$, $P = 1.4 \times 10^{-57}$),

\begin{equation}
	\log \tau_{\rm DRW}^{\rm in} = (0.50\pm0.02)\log M_{\rm BH}+(-2.03\pm0.19),
\label{tauinjet}
\end{equation}        
with a $1\sigma$ intrinsic scatter of $0.14$ dex. \cite{Kelly2009} investigated the relation between the optical variability timescale and black hole mass for 70 quasars and found that $\tau_{\rm{DRW}} \propto M_{\rm BH}^{0.56\pm0.14}$, indicating a power-law dependence on black hole mass. Using SDSS Stripe 82 quasars, \cite{MacLeod2010} reported a similar scaling: $\tau_{\rm{DRW}} \propto M_{\rm BH}^{0.14\pm0.12}\lambda_{\rm Edd}^{-0.08\pm0.10}$. \cite{Kozlowski2016} analyzed 518 AGNs and found $\tau_{\rm{DRW}} \propto M_{\rm BH}^{0.38\pm0.15}$, confirming a significant correlation with black hole mass. \cite{Guo2017}, based on the SDSS Stripe 82 Quasar Sample, derived $\tau_{\rm{DRW}} \propto M_{\rm BH}^{0.62}\lambda_{\rm Edd}^{0.48}$, highlighting the additional role of Eddington ratio. For 67 non-jetted systems, \cite{Burke2021} obtained a slope of $\log \tau_{\rm{DRW}} \sim (0.38\pm0.05)\log M_{\rm BH}$. \cite{Arevalo2024} also reported $\tau_{\rm{DRW}} \propto M_{\rm BH}^{0.62}\lambda_{\rm Edd}^{0.48}$, consistent with earlier findings. In a study of 79 dwarf AGNs, \cite{Wang2023} found $\tau_{\rm{DRW}} \propto M_{\rm BH}^{0.31\pm0.03}$. \cite{Su2024} identified a scaling relation $\tau_{\rm{DRW}} \propto M_{\rm BH}^{0.6-0.8}$, while \cite{Zhang2024} measured $\log \tau_{\rm{DRW}} \sim (0.57\pm0.02)\log M_{\rm BH}$ for 34 blazars and 7 microquasars. Our results are consistent with these previous studies within uncertainties. Meanwhile, we found that the slopes of Equation \ref{tauinall} and Equation \ref{tauinjet} are similar within the error range. These results may suggest that jetted and non-jetted AGNs may share similar physical properties. \cite{Scaringi2015} and \cite{Burke2021} identified an optical-band scaling relation between non-jetted stellar-mass black holes and supermassive black holes, indicating that black hole accretion systems across different mass scales have the same physical nature. Notably, the scaling relation between the optical variability timescale and black hole mass in our sample agrees well with the prediction from the standard thin disk model \citep{Burke2021}, i.e., $\tau_{\rm{DRW}} \propto M_{\rm BH}^{0.50}$, suggesting that accretion disk variability may influence jet emission or be linked to physical processes within the jet.

Figure~\ref{tauMBH} shows the best-fitting correlation between the optical variability timescale and black hole mass for low-mass microquasars and high-mass AGNs. This observation raises the question of whether relativistic jet emissions are influenced by the host black hole mass. Specifically, are the emission mechanisms within the jets or the processes driving jet formation independent of black hole mass? These questions are fundamental and warrant systematic investigation. \cite{Liodakis2017} identified a strong correlation between the intrinsic broadband radio luminosity of blazars and black hole mass, extending this relationship to stellar-mass black hole systems such as microquasars. This result places significant constraints on alternative jet models and suggests that jet production mechanisms are universal across black hole systems, regardless of mass. Furthermore, the study indicates that AGNs operate in a similar accretion regime as the hard state observed in microquasars. Our findings also provide indirect evidence that optical-band variability—driven by physical processes within relativistic jets—may be a universal phenomenon across black hole systems, independent of their mass. Although jets in microquasars extend over shorter spatial scales compared to those in AGNs, they are sufficiently efficient to produce analogous variability patterns. The scaling relation obtained in this work also stand out as an independent method for estimating the Doppler factor.  Estimates of black hole mass and the optical variability timescale can be applied to determine the Doppler factor within the dispersion of the best-fit correlation. 

\begin{figure}
	\includegraphics[width=8.0cm,height=5.0cm]{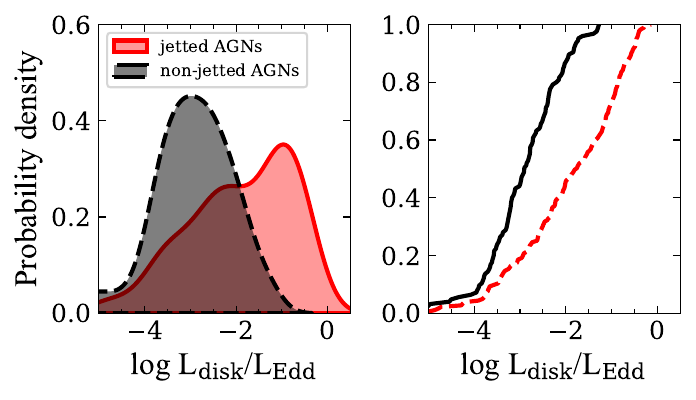}
	\centering
	\caption{The distribution of accretion rates (left) and its cumulative distribution (right) for jetted and non-jetted AGNs. The red line is jetted AGNs, and black line is non-jetted AGNs.}
	\label{histaccretion}
\end{figure}

In addition to the beaming effect, we consider the difference in accretion rates for jetted and non-jetted AGNs. The distribution of accretion rates for jetted and non-jetted AGNs is shown in Figure.\ref{histaccretion}. The average accretion rates for jetted and non-jetted AGNs are $\overline{<\log L_{\rm disk}/L_{\rm Edd}>}|_{\rm jetted}=-1.90\pm1.12$ and $\overline{<\log L_{\rm disk}/L_{\rm Edd}>}|_{\rm jetted}=-2.91\pm0.81$, respectively. We find that jetted AGNs have a higher average accretion rate compared to non-jetted AGNs. Through the Kolmogorov-Smirnov (KS) test, we find a significant difference in the distribution of accretion rates between jetted and non-jetted AGNs ($p=1.2\times10^{-13}$). The non-jetted AGNs typically have lower accretion rates and lower black hole spins compared to jetted AGNs, which are driven by more efficient accretion processes. Jetted AGNs often feature more massive black holes and higher accretion rates, producing powerful jets, whereas non-jetted AGNs are dominated by thermal emission from the disk \citep{Ballantyne2007}.            

\begin{figure*}
	\includegraphics[width=16.0cm,height=8.0cm]{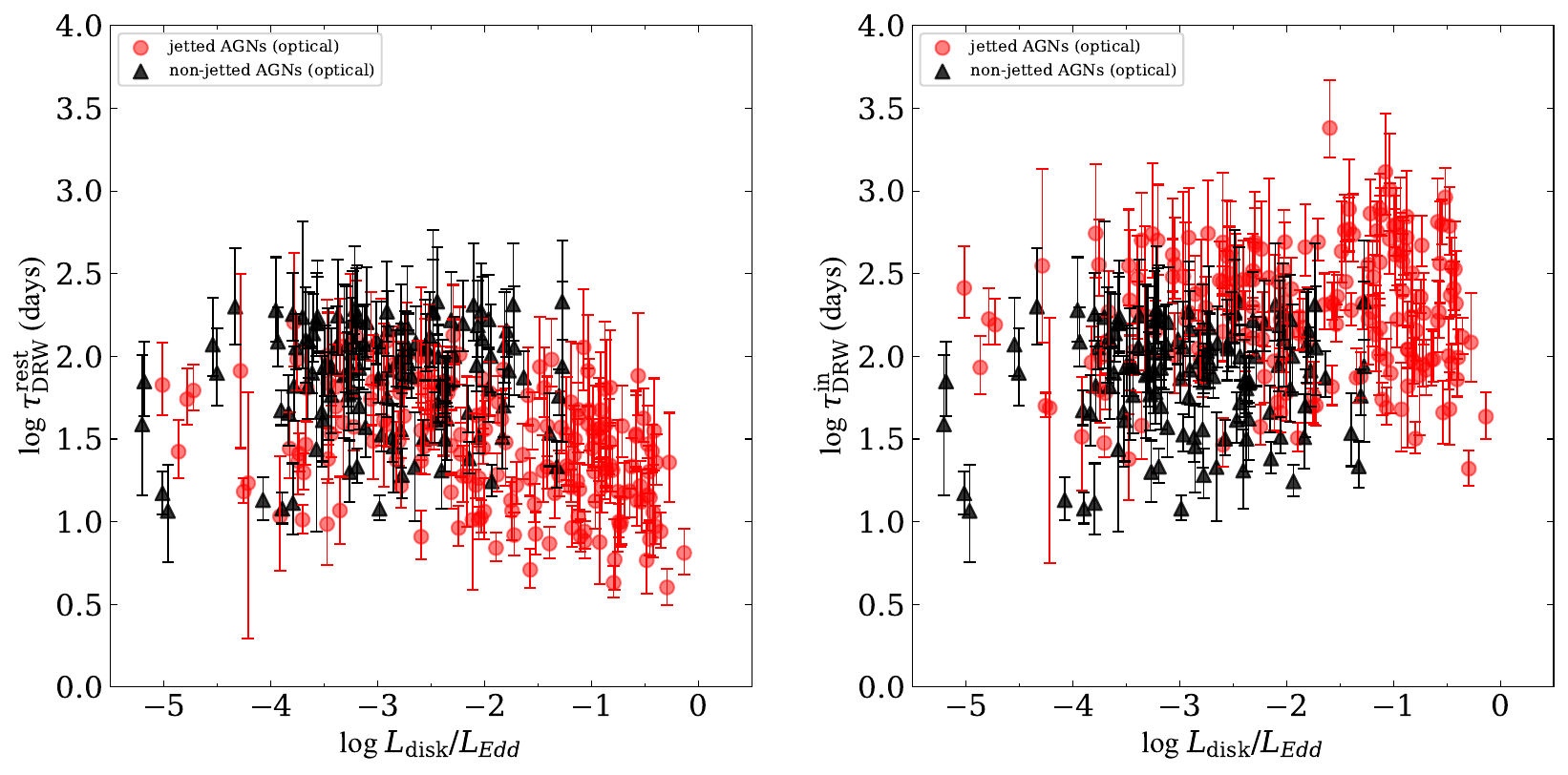}
	\centering
	\caption{Relation between the rest-frame optical variability timescale (left panel) and the intrinsic optical variability timescale (right panel) and accretion rates for jetted and non-jetted AGNs. The red dot is jetted AGNs, and black dot is non-jetted AGNs.}
	\label{tauaccretion}
\end{figure*}

The relation between optical variability timescale and accretion rates for jetted and non-jetted AGNs is shown in Figure.\ref{tauaccretion}. From the left panel of figure.\ref{tauaccretion}, we find that there is a significant correlation between the rest-frame optical variability timescale and accretion rates for jetted AGNs ($r=-0.47, p=6.5\times10^{-8}$). However, there is no correlation between the rest-frame optical variability timescale and accretion rates for non-jetted AGNs ($r=0.11, p=0.23$). At the same time, we find that, once the effects of relativistic Doppler boosting are corrected for, the optical variability timescale of jetted AGNs is also not correlated with the accretion rate ($r=0.087, p=0.24$). \cite{Kelly2009} found that there is no correlation between the optical variability timescale and the accretion rate for 70 quasars. \cite{Lu2019} found that there is no correlation between the optical variability timescale and the accretion rate for 73 AGNs. \cite{Zhang2022} studied the relationship between the gamma-ray variability timescale and accretion rates for 15 AGNs. They found that the gamma-ray variability timescale is independent of the accretion rates for 15 AGNs. Our results are consistent with theirs.          

\section{Conclusions}
We present the optical variability of 125 non-jetted AGNs based on ZTF light curves. To characterize this variability, we model the light curves using a damped random walk (DRW) process. Our main findings are summarized as follows.

(1) Our results show that the characteristic variability timescales of both jetted and non-jetted AGNs are in close agreement, regardless of differences in black hole mass.

(2) The similarity in variability timescales observed in stellar-mass black holes and supermassive black holes suggests that jets may be produced through a common physical mechanism.

(3) We confirm the correlation between the rest-frame DRW variability timescale and black hole mass by combining our sample of 125 non-jetted and 183 jetted AGNs with data from \cite{Burke2021}, \cite{Wang2023}, and \cite{Su2024}. A linear fit is applied to the combined dataset of 464 sources, yielding a slope of $\sim$0.35–0.50, consistent with previous studies.

\section*{Acknowledgements}
Yongyun Chen is grateful for financial support from the National Natural Science Foundation of China (No. 12203028). This work was support from the research project of Qujing Normal University (2105098001/094). This work is supported by the youth project of Yunnan Provincial Science and Technology Department (202101AU070146, 2103010006). Yongyun Chen is grateful for funding for the training Program for talents in Xingdian, Yunnan Province. QSGU is supported by  the National Natural Science Foundation of China (No. 12192222, 12192220 and 12121003). We also acknowledge the science research grants from the China Manned Space Project with NO. CMS-CSST-2021-A05. This work is supported by the National Natural Science Foundation of China (11733001 and U2031201). D.R.X. acknowledge the science research grants from the China Manned Space Project with No. CMS-CSST- 2021-A06, Yunnan Province Youth Top Talent Project (YNWR-QNBJ-2020-116) and the CAS “Light of West China” Program. Xiaogu Zhong acknowledges the financial supports from the Yunnan Fundamental Research Projects (grant NO. 202501AT070392) and the Science Foundation of Department of Education of Yunnan Province (2024J0935).

\section*{Data Availability}
All the data used here are available upon reasonable request. All datas are in Table 1. 

\bibliographystyle{mnras}
\bibliography{example} % if your bibtex file is called example.bib

%\begin{thebibliography}{99}
%\end{thebibliography}

%%%%%%%%%%%%%%%%%%%%%%%%%%%%%%%%%%%%%%%%%%%%%%%%%%

%%%%%%%%%%%%%%%%% APPENDICES %%%%%%%%%%%%%%%%%%%%%
% Don't change these lines
\bsp	% typesetting comment
\label{lastpage}
\end{document}